\documentclass[aps,pra,twocolumn,showpacs,superscriptaddress,10pt]{revtex4-2}

\usepackage[colorlinks ,linkcolor=magenta,anchorcolor=green,citecolor=blue,urlcolor=blue]{hyperref}
\usepackage{amsmath,amsfonts,amssymb,mathrsfs}
\usepackage{bm,bbm}
\usepackage{graphicx}
\usepackage[dvipsnames]{xcolor}
\usepackage{verbatim}
\usepackage[normalem]{ulem}

\usepackage{physics}

\usepackage[T1]{fontenc}
\usepackage{tgtermes}

\usepackage{hyperref}
\usepackage{xurl}
\hypersetup{breaklinks=true}

\newcommand{\iu}{\mathrm{i}\mkern1mu}
\newcommand{\fG}{f\mathcal{G}\mkern1mu}
\newcommand{\G}{\mathcal{G}\mkern1mu}
\newcommand{\CZ}{\mathrm{CZ}\mkern1mu}
\newcommand{\CCZ}{\text{c-}\widehat{\CZ}\mkern1mu}
\newcommand{\CfG}{\text{c-}\widehat{\fG}\mkern1mu}

\begin{document}

\title{Preparing the Gutzwiller wave function for attractive SU(3) fermions on a quantum computer}

\author{Han Xu}
\affiliation{Computational Materials Science Research Team, RIKEN Center for Computational Science (R-CCS), Hyogo 650-0047, Japan}
\email{han.xu@riken.jp}

\author{Kazuhiro Seki}
\affiliation{Quantum Computational Science Research Team, RIKEN Center for Quantum Computing (RQC), Saitama 351-0198, Japan}

\author{Seiji Yunoki}
\affiliation{Computational Materials Science Research Team, RIKEN Center for Computational Science (R-CCS), Hyogo 650-0047, Japan}
\affiliation{Quantum Computational Science Research Team, RIKEN Center for Quantum Computing (RQC), Saitama 351-0198, Japan}
\affiliation{Computational Quantum Matter Research Team, RIKEN Center for Emergent Matter Science (CEMS), Saitama 351-0198, Japan}
\affiliation{Computational Condensed Matter Physics Laboratory, RIKEN Pioneering Research Institute (PRI), Saitama 351-0198, Japan}

\begin{abstract}
    We implement the Gutzwiller wave function for attractive SU(3) fermion systems on a quantum computer using a quantum-classical hybrid scheme based on the discrete Hubbard-Stratonovich transformation.
    In this approach, the nonunitary Gutzwiller operator is decomposed into a linear combination of unitaries constructed from two-qubit fermionic Givens rotation gates, whose rotation angles are dictated by the auxiliary fields. 
    We develop and reformulate two complementary methods to perform the sum over these auxiliary fields.
    In the first method, the Gutzwiller wave function is probabilistically prepared on the register qubits by projectively postselecting the desired state via measurements of ancilla qubits. 
    We analyze the success rate both analytically and numerically as a function of the Gutzwiller variational parameter $g$ for the Fermi-sea and BCS-like trial states at half filling. The success rate is found to decay exponentially for small $|g|$, but remains finite in the $|g|\to\infty$ limit, with increasing $|g|$. 
    In the second method, we employ importance sampling to address the Gutzwiller variational problem, where the central objective is to estimate the expectation values of observables.  
    We demonstrate the proposed scheme by calculating the energy and triple occupancy of the attractive SU(3) Hubbard model in the framework of digital quantum simulation.
    Moreover, we present experimental results obtained on a trapped-ion quantum computer for the two-site attractive SU(3) Hubbard model, showing good agreement with exact values within statistical errors. 
\end{abstract}

\date{\today}

\maketitle

\section{Introduction}\label{sec:intro}

Quantum computers can naturally access the exponentially large Hilbert spaces that arise in quantum many-body systems, which typically demand exponential computational resources on classical computers~\cite{feynman1982,benioff1980computer}.
With advances in quantum simulation algorithms~\cite{Ortiz2001Quantum,Wecker2015Progress,Wecker2015Solving,Jiang2018Quantum,Malley2016Scalable,kandala2017hardware}, quantum computing has attracted growing interest as a promising platform for studying strongly correlated quantum systems. 
In particular, the variational quantum eigensolver~\cite{peruzzo2014variational} and quantum imaginary time evolution~\cite{mcardle2019variational,motta2020determining} have emerged as prominent quantum algorithms for obtaining ground and excited states. 
The success of these variational approaches critically depends on the choice of the initial state, highlighting the importance of physically motivated state preparation schemes~\cite{Murta2021Gutzwiller}.
A conceptually simple yet physically meaningful candidate is the Gutzwiller wave function~\cite{Gutzwiller1963,Gutzwiller1965,Bunemann1998}, which has been widely used in condensed matter physics. It introduces a single  variational parameter that effectively captures electron correlation effects on top of a trial state, such as a single Slater-determinant state~\cite{Gutzwiller1963}.

Gutzwiller-type wave functions for two-component fermions have been widely used to describe the ground states of various strongly correlated electron systems, including the dimer model~\cite{Fabrizio2007Gutzwiller}, the Haldane-Shastry model~\cite{Haldane1988Exact,Shastry1988Exact}, and $t$-$J$-type models~\cite{yokoyama1987hubbard,Kuramoto1991Exactly,Yokoyama1991Variational,Himeda2000Spontaneous,Himeda2002Stripe,yunoki2005coherent,yunoki2006single,Lee2006Doping}. In classical computation, the Gutzwiller wave function can be straightforwardly implemented by applying a basis transformation in the full Hilbert space. In contrast, implementing a Gutzwiller-type wave function on a quantum computer is significantly more challenging due to the nonunitarity nature of the Gutzwiller operator~\cite{Mazzola2019Nonunitary}. 
Nevertheless, several schemes have been proposed to realize the Gutzwiller operator on a quantum computer~\cite{Mazzola2019Nonunitary,Murta2021Gutzwiller,Yao2021Gutzwiller,Seki2022Gutzwiller,Stenger2023Implementing}.
For instance, Mazzola {\it et al}. employed a truncated expansion of the nonunitary factor to evaluate energy expectation values~\cite{Mazzola2019Nonunitary}. 
Yao {\it et al.} utilized an embedding approach that maps the interacting system to a noninteracting quasiparticle model, enabling variational energy minimization within the variational quantum eigensolver framework~\cite{Yao2021Gutzwiller}. 
Murta {\it et al.} designed a quantum circuit incorporating ancillary qubits to probabilistically prepare the Gutzwiller wave function~\cite{Murta2021Gutzwiller}. 
More recently, Seki {\it et al.} proposed a quantum-classical hybrid framework based on the discrete Hubbard-Stratonovich (HS) transformation, in which the Gutzwiller operator is represented as a linear combination of unitaries at the cost of introducing auxiliary fields~\cite{Seki2022Gutzwiller}.

Here, we extend the discrete HS transformation-based scheme~\cite{Seki2022Gutzwiller} to systems of interacting SU(3) fermions and demonstrate its implementation on a quantum computer. 
Interacting SU(3) fermions have attracted significant attention across diverse areas of physics, from high energy physics~\cite{Fodor2002,Aoki2006,Wilczek2007} to condensed matter systems and ultracold atomic gases~\cite{Honerkamp2004Ultra,Honerkamp2004BCS}. Specifically, ultracold fermionic $^6\mathrm{Li}$ atoms confined in the three lowest hyperfine states can experimentally realize SU(3) symmetry when the $s$-wave scattering lengths are tuned to the same negative value~\cite{Abraham1997,Bartenstein2005,Ottenstein2008,Huckans2009}.
Three-component interacting fermion systems exhibit phenomena that go beyond those observed in conventional electronic systems, such as color superfluid and trionic states, as described in the attractive SU(3) Hubbard model~\cite{Rapp2007Color,Rapp2008Trionic,Inaba2009,inaba2011color,Titvinidze2011,Koga2017,Xu2023Trion,Li2023Quantum,chetcuti2023probe}.
In this context, the Gutzwiller wave function has been employed to investigate quantum phase transitions in the intermediate coupling regime, where perturbative methods fail~\cite{Rapp2007Color,Rapp2008Trionic}.

More precisely, in this study, we extend the previous work in 
Ref.~\cite{Seki2022Gutzwiller} to prepare the Gutzwiller wave function 
for the attractive SU(3) fermion Hubbard model on a quantum computer. 
Our approach enables the representation of the nonunitary 
Gutzwiller operator as a linear combination of unitaries, allowing 
its implementation via quantum circuits constructed from fermionic 
Givens rotation gates.  
We develop two complementary methods: a probabilistic state 
preparation using ancilla-based postselection, and a measurement-based 
scheme leveraging importance sampling without the use of ancillary 
qubit. 
These methods are validated through numerical simulations of the 
attractive SU(3) Hubbard model on one-dimensional chains and square 
lattices, and further demonstrated through a proof-of-principle 
implementation on a trapped-ion quantum computer for a two-site 
system. 
Our results provide a practical framework 
for implementing Gutzwiller-type variational states for 
multi-component fermionic systems on quantum hardware.

The remainder of this paper is organized as follows.  
In Sec.~\ref{sec:model}, we introduce the attractive SU(3) Hubbard 
model and define the Gutzwiller wave function and Gutzwiller operator. 
We also formulate the Gutzwiller variational problem for a two-site 
system as a pedagogical example.
We then present the discrete Hubbard–Stratonovich transformation 
tailored for the SU(3) interaction, and outline the Jordan–Wigner 
transformation used to map fermionic operators onto qubits.
In Sec.~\ref{sec:implementation}, we detail two complementary 
implementation strategies based on a linear-combination-of-unitaries 
circuit and an importance-sampling-based evaluation of observables. 
We highlight the conceptual differences between the present scheme 
and the SU(2) case studied in Ref.~\cite{Seki2022Gutzwiller}, 
emphasizing the distinct structure of the SU(3) Hubbard interaction.
Numerical results for small clusters are provided for both 
one-dimensional and two-dimensional lattices.
We then focus on a practical realization of our circuit for 
a two-site system and execute it on both ideal and noisy quantum 
simulations, as well as on a trapped-ion quantum device.
In Sec.~\ref{sec:conc}, we summarize our findings and discuss 
prospects for scaling up the approach to larger lattice systems.


\section{Model and formalism}\label{sec:model}

\subsection{Attractive SU(3) Hubbard model}

We consider the attractive SU(3) Hubbard model defined on a lattice 
by the Hamiltonian:
\begin{equation}\label{eq:hamiltonian}
    \hat H = \hat K + U\hat D,
\end{equation}
where
\begin{equation}
    \hat K = -J\sum_{\langle i,j\rangle}\sum_{\alpha=1}^3(\hat{c}_{i\alpha}^{\dagger}\hat{c}_{j\alpha}+\text{H.c.})
\end{equation}
and
\begin{equation}
    \hat D = \sum_{i=1}^{N_{\text{site}}}\sum_{\alpha<\beta}\left(\hat{n}_{i\alpha}-\frac{1}{2}\right)\left(\hat{n}_{i\beta}-\frac{1}{2}\right).
\end{equation}
Here, $\langle{i,j}\rangle$ denotes nearest-neighbor pairs of sites 
and $\alpha,\beta\in\{1,2,3\}$ are SU(3) color indices. 
The operator $\hat c_{i\alpha}^\dag$ ($\hat c_{i\alpha}$) creates 
(annihilates) a fermion of color $\alpha$ at site $i$, 
and $\hat{n}_{i\alpha}=\hat{c}_{i\alpha}^{\dagger}\hat{c}_{i\alpha}$ 
is the corresponding number operator. 
The parameter $U<0$ denotes the on-site attractive interaction between 
fermions of different colors. 
We set the nearest-neighbor hopping amplitude $J=1$ as the energy unit 
throughout this work. The total number of lattice sites is denoted by 
$N_{\rm site}$.

\subsection{Gutzwiller wave function}

For the attractive SU(3) Hubbard model, the Gutzwiller wave function 
was introduced in Ref.~\cite{Rapp2007Color} and is defined as 
\begin{equation}
    \ket{G} = \hat{P}(\tilde{g})\ket{\psi_0}
    = \prod_{i=1}^{N_{\text{site}}}\left[1+(\tilde{g}-1)\hat{n}_{i1}\hat{n}_{i2}\hat{n}_{i3}\right]\ket{\psi_0},
    \label{eq:G}
\end{equation}
where $\hat{P}(\tilde{g})$ is the Gutzwiller operator, and 
$\tilde{g}$ ($\geqslant1$) is a variational parameter that 
enhances the weight of triply occupied sites. When $\tilde{g}=1$, 
the Gutzwiller wave function reduces to the trial state 
$\ket{\psi_0}$, which is typically chosen to be a Fermi-sea or 
BCS-like state. 
In the limit $\tilde{g}\to\infty$, the projection strongly 
favors configurations with triply occupied sites, and 
$\hat{P}(\tilde{g})\ket{\psi_0}$ approaches a superposition of 
trionic states~\cite{Rapp2008Trionic}.

To express the Gutzwiller wave function in terms of density-density 
interaction operators, we introduce an alternative Gutzwiller 
operator defined as 
\begin{equation}
    \hat G(\tilde{g}) = \prod_{i=1}^{N_{\text{site}}}\prod_{\alpha<\beta}(1+(\tilde{g}-1)\hat{n}_{i\alpha}\hat{n}_{i\beta}), 
    \label{eq:gw0}
\end{equation}
which is essentially equivalent to 
the operator introduced in Eq.~(\ref{eq:G}), apart from 
additional terms proportional to 
$\hat{n}_{i1}\hat{n}_{i2}+\hat{n}_{i2}\hat{n}_{i3}+\hat{n}_{i3}\hat{n}_{i1}$.
Taking normalization into account, we define the normalized 
Gutzwiller wave function as a variational state: 
\begin{equation}\label{eq:gutzwiller-exp-p}
    \ket{{\tilde\psi}_{\tilde{g}}} = \frac{\hat G(\tilde{g})\ket{\psi_0}}{\sqrt{\expval{\hat G(\tilde{g})^2}{\psi_0}}},
\end{equation}
where $\ket{\psi_0}$ is a normalized trial state.
As with the wave function $\ket{G}$ defined in Eq.~\eqref{eq:G}, 
the state $\ket{\psi_{\tilde{g}}}$ reduces to the trial state 
$\ket{\psi_0}$ when $\tilde{g}=1$, and approaches 
a superposition of trionic states in the limit 
$\tilde{g}\to \infty$. It should be noted, however, that 
in general $\hat{P}(\tilde{g}) \not = \hat{G}(\tilde{g})$.


Noting that 
\begin{eqnarray}
e^{-g\hat D} = e^{-\frac{3g}{4}N_{\rm site}}
e^{g\sum_{i,\alpha}\hat{n}_{i\alpha}}
\prod_{i=1}^{N_{\text{site}}}\prod_{\alpha<\beta}e^{-g\hat{n}_{i\alpha}\hat{n}_{i\beta}} 
\end{eqnarray}
and 
\begin{eqnarray}
e^{-g\hat{n}_{i\alpha}\hat{n}_{i\beta}}=1+(e^{-g}-1)\hat{n}_{i\alpha}\hat{n}_{i\beta}
\end{eqnarray}
for $\alpha\ne\beta$, we find that 
\begin{equation}
\hat G(\tilde{g})\ket{\psi_0} \propto e^{-g\hat D}\ket{\psi_0},
\end{equation}
as long as the total particle number is conserved and 
$g = -\log{\tilde{g}}$ ($\leqslant0$). 
%
The operator $e^{-g\hat D}$ is particularly suitable for 
implementation via the discrete HS transformation, as will be 
discussed below. Therefore, 
in what follows, we adopt $e^{-g\hat D}$ as the Gutzwiller operator 
for the attractive SU(3) Hubbard model and define the Gutzwiller 
wave function as  
\begin{equation}\label{eq:gutzwiller-exp}
    \ket{\psi_g} = \frac{e^{-g\hat D}\ket{\psi_0}}{\sqrt{\expval{e^{-2g\hat D}}{\psi_0}}},
\end{equation}
even when the trial state $|{\psi_0}\rangle$ is not an eigenstate 
of the total particle number operator, such as the BCS-like state 
introduced in the next subsection.

\subsection{Gutzwiller problem for the two-site model} 
\label{sec:model:twosite}

We now demonstrate the Gutzwiller variational procedure by 
explicitly deriving the variational energy for the two-site 
attractive SU(3) Hubbard model at half filling.

We begin by considering the Fermi-sea trial state, i.e., the ground 
state of $\hat K$: 
\begin{equation}
    \ket{\psi_0} = \prod_{\alpha=1}^3 \frac{1}{\sqrt{2}}(\hat c_{1\alpha}^{\dagger}+\hat c_{2\alpha}^{\dagger})\ket{\rm vac},
\end{equation}
where $\ket{\rm vac}$ denotes the fermionic vacuum, satisfying 
$\hat c_{i\alpha}\ket{\rm vac}=0$ for all $i$ and $\alpha$.
Applying the Gutzwiller operator to $\ket{\psi_0}$ yields 
\begin{equation}
    e^{-g\hat D}\ket{\psi_0} = \sqrt{\frac{1}{4}}e^{-\frac{3}{2}g}\ket{T} + \sqrt{\frac{3}{4}}e^{\frac{1}{2}g}\ket{D},
\end{equation}
where $\ket{T}$ is the normalized superposition of the two triply 
occupied states and $\ket{D}$ is that of the six doubly occupied 
states. The action of $\hat K$ on this state is given by 
\begin{align}
        \hat Ke^{-g\hat D}\ket{\psi_0}=&-3J\sqrt{\frac{1}{4}}e^{\frac{1}{2}g}\ket{T} \nonumber \\
        &-J\sqrt{\frac{3}{4}}(e^{-\frac{3}{2}g}+2e^{\frac{1}{2}g})\ket{D}        
\end{align} 
Thus, the expectation value of the kinetic term 
(up to normalization) is computed as
\begin{equation}
    \expval{e^{-g\hat D}\hat Ke^{-g\hat D}}{\psi_0}=-3J\cosh(g).
\end{equation}
The expectation value of the interaction term (up to 
normalization) can be computed using 
the identity 
\begin{equation}
    \begin{aligned}
        \expval{e^{-g\hat D}U\hat De^{-g\hat D}}{\psi_0}=&-\frac{1}{2}\partial_{g}\expval{e^{-2g\hat D}}{\psi_0} \\
        =&-\frac{1}{2}\partial_{g}(\frac{1}{4}e^{-3g}+\frac{3}{4}e^{g}) \\
        =&-\frac{3}{4}Ue^{-g}\sinh(2g),
    \end{aligned}
\end{equation}
where $\partial_g$ denotes the partial derivative with respect to 
$g$, i.e., $\partial/\partial g$.

Therefore, the variational energy is given by
\begin{equation}\label{eq:opt-energy}
    \begin{aligned}
        E(g) &= \expval{\hat K}{\psi_g} + U\expval{\hat D}{\psi_g} \\
        &= \frac{-12J\cosh(g)-3Ue^{-g}\sinh(2g)}{e^{-3g}+3e^{g}}.
    \end{aligned}
\end{equation}
Solving the stationary condition 
$\partial_g E(g)|_{g=g_{\text{opt}}}=0$ yields the optimal 
variational parameter, 
\begin{equation}\label{eq:opt-g}
    e^{2g_{\text{opt}}} = \frac{1}{3J}(U+J+\sqrt{(U+J)^2+3J^2}).
\end{equation}
Substituting Eq.~\eqref{eq:opt-g} into Eq.~\eqref{eq:opt-energy} 
gives the optimized variational energy:  
\begin{equation}
    E(g_{\text{opt}}) = \frac{1}{2}(U-2J-2\sqrt{(U+J)^2+3J^2}),
\end{equation}
which exactly coincides with the ground-state energy of the two-site 
attractive SU(3) Hubbard model at half filling.

Another choice of the trial state is the BCS-like state, 
defined as~\cite{Rapp2007Color,Rapp2008Trionic}
\begin{equation}\label{eq:bcs-trial}
    \ket{\psi_0} = \prod_{k<k_F}\hat c_{k3}^{\dagger}\prod_{k'}(u_{k'}+v_{k'}\hat c_{k'1}^{\dagger}\hat c_{-k'2}^{\dagger})\ket{\rm vac},
\end{equation} 
where $u_k^2=\frac{1}{2}(1+\epsilon_k/\sqrt{\epsilon_k^2+\Delta^2})$ 
and $v_k^2=1-u_k^2$ are the BCS coherence factors,
with $\epsilon_{k}=-J\cos{k}$ and $\Delta$ being the BCS gap 
parameter. Here, we implicitly assume that the chemical potential is 
zero. 
Using the Fourier transformation $c_{k\alpha}^{\dagger}=\frac{1}{\sqrt{N_{\text{site}}}}\sum_{i} c_{i\alpha}^{\dagger}e^{ikr_i}$ 
for each fermion color, the fermion creation operators at wave 
vectors $k_1=0$ and $k_2=\pi$ are given by 
$c_{k_1,\alpha}^{\dagger}=\frac{1}{\sqrt{2}}(c_{1\alpha}^{\dagger}+c_{2\alpha}^{\dagger})$ 
and $c_{k_2,\alpha}^{\dagger}=\frac{1}{\sqrt{2}}(c_{1\alpha}^{\dagger}-c_{2\alpha}^{\dagger})$, respectively. 
Note also that $u_{k_2}^2 = v_{k_1}^2$, $u_{k_1}^2 = v_{k_2}^2$, 
and $u_{k_1}v_{k_1} = u_{k_2}v_{k_2}$. 
Since the explicit form of the Gutzwiller wave function 
$\ket{\psi_g}$ 
becomes quite cumbersome in this case, we present only the 
expectation values of the Hamiltonian with respect to 
$e^{-g\hat D}\ket{\psi_g}$:
\begin{equation}
    \begin{aligned}
        &\expval{e^{-g\hat D}\hat Ke^{-g\hat D}}{\psi_0} = \\ 
    &- 3J \cosh(g)  + 4Ju^2[\cosh(g)+v^2\sinh(g)],
    \end{aligned}
\end{equation}
and
\begin{equation}\label{eq:bcs-d}
    \begin{aligned}
    &\expval{e^{-g\hat D}U\hat De^{-g\hat D}}{\psi_0} = \\
    &-\frac{3}{4} Ue^{-g}\sinh(2g) + 2u^2 v^2 U\cosh(g),
    \end{aligned}
\end{equation}
where $u= u_{k_1}$ and $v= v_{k_1}$. In addition, the normalization 
factor is given by 
\begin{equation}
    \expval{e^{-2g\hat D}}{\psi_0} = 
    \frac{1}{4}(e^{-3g} + 3e^{g}) - 4u^2v^2\sinh(g)
\end{equation}
Thus, the variational energy $E(g,\Delta)$ is obtained for 
the BCS-like trial state. In the limit $\Delta=0$, we recover 
$u=0$ and $v=1$, and the energy expression $E(g,0)$  reduces to 
Eq.~\eqref{eq:opt-energy}, as expected.

\subsection{Discrete Hubbard-Stratonovich transformation}

In the SU(2) Hubbard model, the discrete HS transformation is 
used to decouple the density-density interactions into exponentials 
of fermion number operators~\cite{Assaad2008}. However, 
applying such a transformation to the SU(3) model does not 
eliminate the sign problem in quantum Monte Carlo simulations. 
Instead, we adopt an alternative discrete HS transformation 
scheme~\cite{Xu2023Trion}: 
\begin{equation}\label{eq:Hubbard-Stratonovich}
    e^{-g(\hat n_{i\alpha}-\frac{1}{2})(\hat n_{i\beta}-\frac{1}{2})}\\
    =\gamma\sum_{s_{i\alpha\beta}=\pm 1} e^{s_{i\alpha\beta}\lambda( \hat c_{i\alpha}^{\dagger}\hat c_{i\beta}-\hat c_{i\beta}^{\dagger}\hat c_{i\alpha})},
\end{equation}
where $s_{i\alpha\beta}(=\pm1)$ is a discrete auxiliary field 
coupled to on-site color-flip processes, 
$\gamma = \frac{1}{2}e^{-\frac{g}{4}}$, 
and $\lambda = \arccos{e^{\frac{g}{2}}}$, with the condition $g<0$. 
Using this transformation, the Gutzwiller operator can be rewritten 
as
\begin{equation}\label{eq:HS-GutzF}
    \begin{aligned}
        e^{-g\hat D} &= \gamma^{3N_{\text{site}}}\prod_{i=1}^{N_{\text{site}}}\prod_{\alpha<\beta}\sum_{s_{i\alpha\beta}=\pm1}e^{s_{i\alpha\beta}\lambda \iu\hat\sigma_{i\alpha\beta}^{y}} \\
        &= \gamma^{3N_{\text{site}}}\prod_{i=1}^{N_{\text{site}}}\prod_{\alpha<\beta}(e^{\lambda \iu\hat\sigma_{i\alpha\beta}^{y}}+e^{-\lambda \iu\hat\sigma_{i\alpha\beta}^{y}}),
    \end{aligned}
\end{equation}
where the factor of $3$ arises from the number of unique interaction 
pairs, and the color-flip operator is defined as 
$\iu\hat\sigma_{i\alpha\beta}^{y}=\hat c_{i\alpha}^{\dagger}\hat c_{i\beta} - \hat c_{i\beta}^{\dagger}\hat c_{i\alpha}$. This 
operator can be interpreted as the $y$-component Pauli matrix 
acting in the two-dimensional color subspace $\{\alpha,\beta\}$ at 
site $i$. As a result, the Gutzwiller operator $e^{-g\hat D}$ is 
expressed as a linear combination of unitary operators by summing 
over all auxiliary field configurations 
$\{s_{i12},s_{i13},s_{i23}\}_{i=1}^{N_{\text{site}}}$.

\subsection{Quantum circuit for the Gutzwiller operator}

To simulate fermionic systems on a quantum computer, we map 
fermionic operators to multiqubit operators.
In this paper, we adopt the Jordan-Wigner transformation to encode 
the fermion creation and annihilation operators into qubit operators 
as follows: 
\begin{eqnarray}
    &&\hat c_{i\alpha}^{\dagger} \mapsto \frac{1}{2}(\hat X_{i_{\alpha}}-\iu\hat Y_{i_{\alpha}})\prod_{j<i_{\alpha}}\hat Z_{j}, \label{eq:JWd}\\
    &&\hat c_{i\alpha} \mapsto \frac{1}{2}(\hat X_{i_{\alpha}}+\iu\hat Y_{i_{\alpha}})\prod_{j<i_{\alpha}}\hat Z_{j},
    \label{eq:JW}
\end{eqnarray}
where $\hat X$, $\hat Y$ and $\hat Z$ are Pauli operators. The label 
$i_{\alpha}$ is a qubit index that corresponds to the fermion mode 
labeled by site $i$ and color $\alpha$, and thus ranges from 1 to 
$3N_{\rm site}$ for SU(3) fermions. 
We consider two schemes for assigning qubit indices: one traverses 
the site index before the color index, and the other traverses the 
color index before the site index. These are referred to as 
the color-uniform labeling ($\{i_{1}\}_{i=1}^{N_{\text{site}}},\{i_{2}\}_{i=1}^{N_{\text{site}}},\{i_{3}\}_{i=1}^{N_{\text{site}}}$) 
and the color-alternating labeling 
($\{i_{1},i_{2},i_{3}\}_{i=1}^{N_{\text{site}}}$), respectively~\cite{Seki2022Spatial}.
It is straightforward to verify that both labeling schemes preserve 
the canonical anticommutation relations 
$\{c_{i\alpha}^{\dagger},c_{j\beta}\}=\delta_{ij}\delta_{\alpha\beta}$, $\{c_{i\alpha}^{\dagger},c_{j\beta}^{\dagger}\}=0$, 
and $\{c_{i\alpha},c_{j\beta}\}=0$.

Using the Jordan-Wigner transformation, the fermion number operator 
is encoded as
\begin{equation}\label{eq:JW-ni}
    \hat c_{i\alpha}^{\dagger}\hat c_{i\alpha} \mapsto \frac{1}{4}(\hat X_{i_{\alpha}}-\iu\hat Y_{i_{\alpha}})(\hat X_{i_{\alpha}}+\iu\hat Y_{i_{\alpha}}) = \frac{1}{2}(\hat I-\hat Z_{i_{\alpha}}),
\end{equation}
where $\hat I$ denotes the identity operator. Similarly, the fermion 
hopping operator is mapped to a product of Pauli operators 
(for $i<j$) as  
\begin{equation}
    \begin{aligned}
        \hat c_{i\alpha}^{\dagger}\hat c_{j\alpha}+\text{H.c.} \mapsto& \frac{1}{2}(\iu\hat X_{i_{\alpha}}\hat Y_{j_{\alpha}}-\iu\hat Y_{i_{\alpha}}\hat X_{j_{\alpha}}) \prod_{i_{\alpha}\leqslant l<j_{\alpha}}\hat Z_{l}, \\
        =& \frac{1}{2}(\hat Y_{i_{\alpha}}\hat Y_{j_{\alpha}}+\hat X_{i_{\alpha}}\hat X_{j_{\alpha}}) \hat Z_{\text{JW},i_{\alpha}j_{\alpha}},
    \end{aligned}
\end{equation}
where we have used the anticommutation relations for Pauli matrices, 
such as $\{\hat Z_{i_{\alpha}},\hat X_{i_{\alpha}}\}=0$, and we 
defined the Jordan-Wigner string 
$\hat Z_{\text{JW},ij}=\prod_{i<l<j}\hat Z_{l}$ 
for notational simplicity. 
Substituting these equations into the Hamiltonian in 
Eq.~\eqref{eq:hamiltonian}, the kinetic and interaction terms 
become 
\begin{equation}\label{eq:JW-K}
    \hat K \mapsto -\frac{J}{2}\sum_{\langle ij\rangle,\alpha}(\hat X_{i_{\alpha}}\hat X_{j_{\alpha}} + \hat Y_{i_{\alpha}}\hat Y_{j_{\alpha}})\hat Z_{\text{JW},i_{\alpha}j_{\alpha}}
\end{equation}
and
\begin{equation}\label{eq:JW-D}
    \hat D \mapsto \frac{1}{4}\sum_{i=1}^{N_{\text{site}}}\sum_{\alpha<\beta}\hat Z_{i_{\alpha}}\hat Z_{i_{\beta}},
\end{equation}
respectively.

\begin{figure}[tb]
    \includegraphics[width=0.96\linewidth]{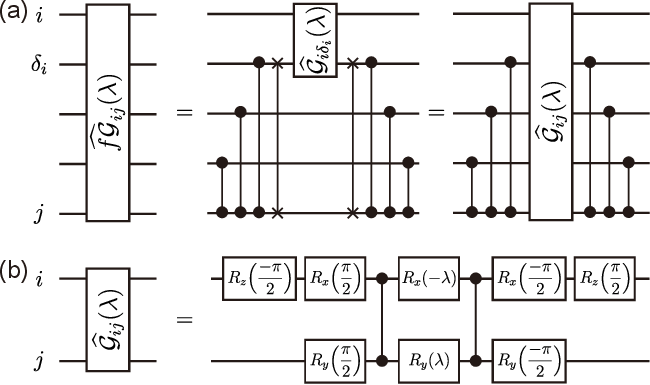}
    \caption{
        (a) Decomposition of the fermionic Givens-rotation gate 
        $\widehat\fG_{ij}(\lambda) $ into a product 
        of CZ gates and a standard Givens-rotation gate. 
        The qubit indicated by $\delta_i$ refers to a qubit 
        located between qubits $i$ and $j$, adjacent to qubit $i$. 
        A solid line connecting two crosses denotes a SWAP gate.
        (b) Decomposition of the Givens-rotation gate 
        $\hat{\cal G}_{ij}(\lambda)$ into two CZ gates and eight 
        single-qubit rotation gates. Here, the single-qubit rotation gate $R_z(\theta)$ is defined as 
        $R_z(\theta) = e^{-\iu\theta\hat Z/2}$, and other rotation 
        gates are defined analogously.}
    \label{fig:Q-rotation}
\end{figure}

On top of this, the color-flip operators 
$\iu\hat\sigma_{i\alpha\beta}^y$ that appear in the HS 
transformation [Eq.~\eqref{eq:Hubbard-Stratonovich}] can be 
mapped via the Jordan-Wigner transformation as (for $\alpha<\beta$) 
\begin{equation}
    \begin{aligned}
        \iu\hat\sigma_{i\alpha\beta}^y \mapsto& \frac{1}{2}(\hat X_{i_{\alpha}}\hat X_{i_{\beta}}+\hat Y_{i_{\alpha}}\hat Y_{i_{\beta}}) \prod_{i_{\alpha}\leqslant l<i_{\beta}}\hat Z_{l}, \\
        =& \iu\frac{1}{2}(\hat X_{i_{\alpha}}\hat Y_{i_{\beta}}-\hat Y_{i_{\alpha}}\hat X_{i_{\beta}}) \hat Z_{\text{JW},i_{\alpha}i_{\beta}}.
    \end{aligned}
\end{equation}
Accordingly, the operators $
e^{\pm\lambda \iu\hat\sigma_{i\alpha\beta}^y}$ in the Gutzwiller 
operator [Eq.~\eqref{eq:HS-GutzF}] can be implemented as 
\begin{equation}\label{eq:given-rotation}
    \begin{aligned}
        e^{\pm\lambda \iu\hat\sigma_{i\alpha\beta}^y} &\mapsto \exp[\iu\frac{\pm\lambda}{2}(\hat X_{i_{\alpha}}\hat Y_{i_{\beta}}-\hat Y_{i_{\alpha}}\hat X_{i_{\beta}}) \hat Z_{\text{JW},i_{\alpha}i_{\beta}}] \\
        &\equiv \widehat\fG_{i_{\alpha}i_{\beta}}(\pm\lambda),
    \end{aligned}
\end{equation}
where $\widehat\fG(\lambda)$ denotes the fermionic Givens-rotation 
gate, as defined in Ref.~\cite{Seki2022Spatial}. 
As shown in Fig.~\ref{fig:Q-rotation}(a), this gate 
$\widehat\fG_{i_{\alpha}i_{\beta}}(\lambda)$ can be decomposed into 
a standard Givens-rotation gate $\widehat\G$ sandwiched between 
controlled-Z (CZ) gates:
\begin{equation}\label{eq:given-rotation-decomposed}
    \widehat\fG_{i_{\alpha}i_{\beta}}(\lambda) = \Big[ \prod_{i_{\alpha}<l<i_{\beta}}\widehat\CZ_{i_{\beta}l} \Big] \widehat\G_{i_{\alpha}i_{\beta}}(\lambda) \Big[ \prod_{i_{\alpha}<l<i_{\beta}}\widehat\CZ_{i_{\beta}l} \Big],
\end{equation}
where 
\begin{equation}
    \widehat\G_{i_{\alpha}i_{\beta}}(\lambda) = \exp[\iu\frac{\lambda}{2}(\hat X_{i_{\alpha}}\hat Y_{i_{\beta}}-\hat Y_{i_{\alpha}}\hat X_{i_{\beta}})]
\end{equation}
is the standard two-qubit Givens-rotation gate. 
In the computational basis ($\ket{00}$,$\ket{01}$,$\ket{10}$,$\ket{11}$), the matrix 
representation of this gate is given by 
\begin{equation}
    \widehat\G_{i_{\alpha}i_{\beta}}(\lambda) = 
    \begin{pmatrix}
        1 & 0 & 0 & 0 \\
        0 & \cos{\lambda} & -\sin{\lambda} & 0 \\
        0 & \sin{\lambda} & \cos{\lambda} & 0 \\
        0 & 0 & 0 & 1
    \end{pmatrix}.
\end{equation}
The Givens-rotation gate $\hat{\cal G}_{i_\alpha,i_\beta}(\lambda)$ 
can be further decomposed into two CZ gates and single-qubit 
rotation gates, as illustrated in Fig.~\ref{fig:Q-rotation}(b).
This decomposition is based on the identity 
\begin{eqnarray}
&&\hat{\cal G}_{i_\alpha,i_\beta}(\lambda)
=
\hat{R}_{z,i_\alpha}(\frac{\pi}{2}) \nonumber \\
&&\times\exp\left[-\iu \frac{\lambda}{2} \left(\hat{X}_{i_\alpha} \hat{X}_{i_\beta} + \hat{Y}_{i_\alpha} \hat{Y}_{i_\beta}\right)\right]
\hat{R}_{z,i_\alpha}(-\frac{\pi}{2})
\end{eqnarray}
together with the standard decomposition of the $XY$-interaction 
unitary  
$\exp\left[-\iu \frac{\lambda}{2} \left(\hat{X}_{i_\alpha} \hat{X}_{i_\beta} + \hat{Y}_{i_\alpha} \hat{Y}_{i_\beta}\right)\right]$
presented in Ref.~\cite{Foss2021Holo}.

Substituting Eq.~\eqref{eq:given-rotation} into 
Eq.~\eqref{eq:HS-GutzF} yields the Gutzwiller operator as 
a linear combination of $2^{3N_{\text{site}}}$ unitary operators: 
\begin{equation}\label{eq:expD-decomposed}
    e^{-g\hat D} \mapsto \sum_{\{s_{i\alpha\beta}\}}\prod_{i=1}^{N_{\text{site}}}\prod_{\alpha<\beta}\gamma \widehat\fG_{i_{\alpha}i_{\beta}}(s_{i\alpha\beta}\lambda),
\end{equation}
where each unitary term is a product of fermionic Givens-rotation 
gates.
From this expression, it becomes evident that using the 
color-alternating labeling, 
$\{i_{1},i_{2},i_{3}\}_{i=1}^{N_{\text{site}}}$, lead to shorter 
Jordan-Wigner strings and fewer CZ gates compared to the 
color-alternating labeling. 
To take advantage of this, we apply a sequence of adjacent 
fermionic SWAP (f-SWAP) gates~\cite{Verstraete2009Quantum} on the 
quantum circuit to switch between color-uniform and 
color-alternating labelings, immediately before and after the 
block that implements the Gutzwiller operator. 
The adjacent swaps required to convert between these labelings 
can be determined via the bubble sort method, resulting in a 
total of  $\frac{3}{2}N_{\text{site}}(N_{\text{site}}-1)$ 
f-SWAP gates in this procedure.

\section{Implementation on a quantum circuit} 
\label{sec:implementation}

\subsection{Approach I: Linear combination of unitary operators} 
\label{sec:implementation-I}

\begin{figure}[tb]
    \includegraphics[width=0.96\linewidth]{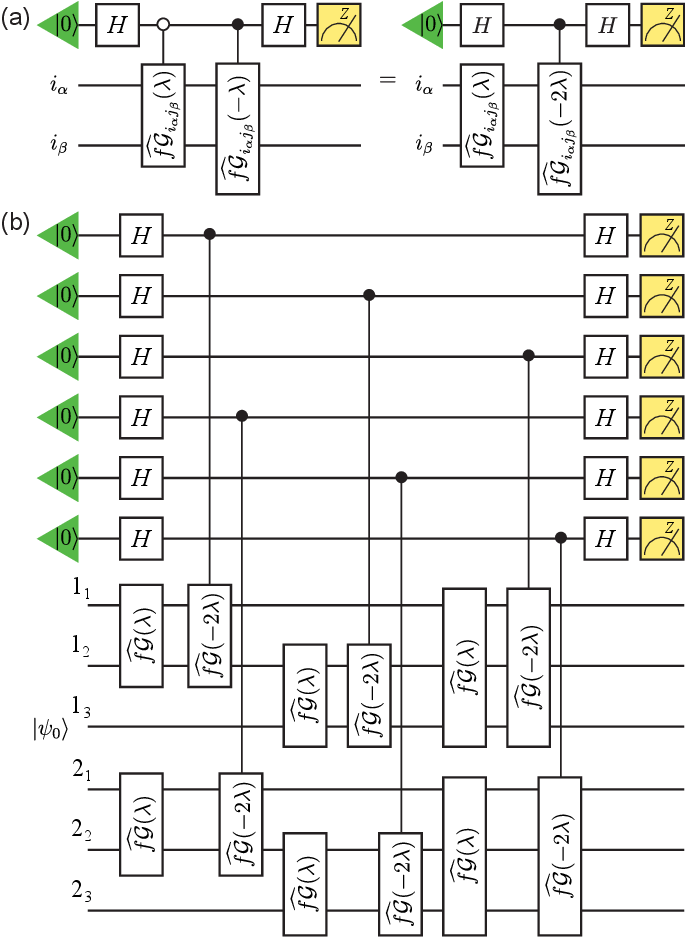}
    \caption{
        (a) Quantum circuits for evaluating the linear combination $\widehat\fG(\lambda)+\widehat\fG(-\lambda)$ using a Hadamard-test-like setup. $H$ denotes the Hadamard gate. Green triangles represent ancillary qubits initialized in the $\ket{0}$ state and the yellow square indicates the measurement in the Pauli basis. 
        (b) Quantum circuit for generating the Gutzwiller wave function $\ket{\psi_g}$ using $3N_{\text{site}}$ ancillary qubits and $3N_{\text{site}}$ register qubits, shown for the case of $N_{\text{site}}=2$. The label $i_{\alpha}$ refers to the qubit encoding the fermion with color $\alpha$ at site $i$, based on the Jordan-Wigner transformation. 
        The ancillary qubits are initialized in the $\ket{0}$ state, while the register qubits are initialized in the trial state $\ket{\psi_0}$, which is prepared using a separate circuit composed of Givens rotation gates.}
    \label{fig:Q-circuit}
\end{figure}

Evaluating Eq.~\eqref{eq:expD-decomposed} requires summing over all auxiliary field configurations, resulting in an exponentially large number of terms in $N_{\text{site}}$. 
Despite this complexity, we present a first approach that directly realizes the linear combination of unitary operators in Eq.~\eqref{eq:expD-decomposed} using $3N_{\text{site}}$ ancillary qubits on a quantum circuit.
Following a similar strategy to that in Ref.~\cite{Seki2022Gutzwiller}, each term of the form $\widehat\fG(\lambda)+\widehat\fG(-\lambda)$ in Eq.~\eqref{eq:expD-decomposed} can be implemented using two consecutive controlled-$\widehat\fG$ ($\CfG$) gates within a Hadamard-test-like quantum circuit, as illustrated in Fig.~\ref{fig:Q-circuit}(a). This construction can be further simplified by replacing the pair of gates with a single $\widehat\fG$ gate followed by a controlled-$\widehat\fG$ gate with double the rotation angle, using the identity $\widehat\fG(\lambda)\widehat\fG(\lambda')=\widehat\fG(\lambda+\lambda')$. 
Using the decomposition of $\widehat\fG$ gate from Fig.~\ref{fig:Q-rotation}(a), the controlled-$\widehat\fG$ gate can be implemented as a controlled-$\widehat\G$ gate sandwiched by products of CZ gates. This makes use of the operator identity for controlled unitary operators: $(\CCZ)(\text{c-}\widehat\G)(\CCZ)=\widehat{\CZ}(\text{c-}\widehat\G)\widehat{\CZ}$ \cite{stair2020multireference}. 
Similarly, the controlled-$\widehat\G$ gate itself requires only two CZ gates, one controlled-$R_x$ (C$R_x$) gate and one controlled-$R_y$ (C$R_y$) gate, based on the decomposition in Fig.~\ref{fig:Q-rotation}(b).
This approach avoids the need for the more complex controlled-controlled-$R_y$ gates that appear in other commonly used gate decompositions~\cite{Jiang2018Quantum}.
In this setting, one ancillary qubit is required for each pair of fermion colors per site, resulting in a total of $3N_{\text{site}}$ ancillary qubits. 
As will be discussed below, this approach is probabilistic in nature.
As a concrete example, Fig.~\ref{fig:Q-circuit}(b) explicitly shows the quantum circuit that implements the Eq.~\eqref{eq:expD-decomposed} for the case of $N_{\text{site}}=2$.

The following provides a detailed analysis of the final quantum state and evaluates the success probability for generating the Gutzwiller wave function on a quantum circuit.
In the Hadamard-test-like setup, each ancillary qubit is initialized in the $\ket{0}$ state. After applying the first Hadamard gate followed by a $\widehat\fG$ gate on the register qubits, the system evolves into the state $\frac{1}{\sqrt{2}}(\ket{0}+\ket{1})\otimes \widehat\fG\ket{\psi_0}$. The subsequent application of the controlled-$\widehat\fG$ gate transforms the system to $\frac{1}{\sqrt{2}}(\ket{0}\otimes \widehat\fG(\lambda)\ket{\psi_0}+\ket{1}\otimes \widehat\fG(-\lambda)\ket{\psi_0})$. Finally, a second Hadamard gate is applied to the ancillary qubit, yielding the state $\frac{1}{2}\ket{0}\otimes (\widehat\fG(\lambda)+\widehat\fG(-\lambda))\ket{\psi_0}+\dots$. Here, the ellipsis represents the components of the wave function where the ancilla is in the state $\ket{1}$. Hence, the desired operator $\prod_{i=1}^{N_{\text{site}}}\prod_{\alpha<\beta}(\widehat\fG_{i_{\alpha}i_{\beta}}(\lambda)+\widehat\fG_{i_{\alpha}i_{\beta}}(-\lambda))$ is successfully applied to the trial state $\ket{\psi_0}$ if and only if all $3N_{\text{site}}$ ancillary qubits are measured in the $\ket{0}$ state.
The final state of the entire system can be expressed as 
\begin{equation}
    \begin{aligned}
        \ket{\Psi} = \ket{0\dots0}\otimes\frac{1}{2^{3N_{\text{site}}}}\prod_{i=1}^{N_{\text{site}}}\prod_{\alpha<\beta}\left(e^{\lambda \iu\hat\sigma_{i\alpha\beta}^{y}}\right. \\
        \left.+e^{-\lambda \iu\hat\sigma_{i\alpha\beta}^{y}}\right)\ket{\psi_0} + \dots,
    \end{aligned}
\end{equation}
where $\ket{0\dots0}$ denotes the product state in which all ancillary qubits are in the $\ket{0}$ state. 
Using Eq.~\eqref{eq:expD-decomposed} and the Born rule, the success probability of generating the Gutzwiller-projected state in the SU(3) case is given by
\begin{equation}\label{eq:p0-expression}
    \begin{aligned}
        p_0 &= \left(\frac{1}{2^{3N_{\text{site}}}\gamma^{3N_{\text{site}}}}\right)^2\expval{e^{-2g\hat D}}{\psi_0} \\
        &= e^{\frac{3}{2}gN_{\text{site}}}\expval{e^{-2g\hat D}}{\psi_0}.        
    \end{aligned}
\end{equation}

Moreover, evaluating the logarithmic derivative $\partial_{|g|}\log{p_0}$ provides further insight into the behavior of the success probability as a function of the site number $N_{\text{site}}$ and the variational parameter $g$.
From Eq.~\eqref{eq:p0-expression}, we obtain
\begin{equation}\label{eq:dp0-expression}
    \begin{aligned}
        \frac{\partial\log{p_0}}{\partial |g| } &= \frac{1}{p_0}\left(
            -\frac{3}{2}N_{\text{site}}p_0
            +p_0\frac{\expval{e^{2|g|\hat D}2\hat D}{\psi_0}}{\expval{e^{2|g|\hat D}}{\psi_0}}\right)\\
            &= -\frac{3}{2}N_{\text{site}} + 2\expval{\hat D}{\psi_g}.
    \end{aligned}
\end{equation}
In the limit $|g|\to0$, the Gutzwiller wave function reduces to the trial state $\ket{\psi_0}$. For the Fermi-sea trial state at half filling, the probability of finding two fermions with different colors on the same site is given by $\expval{n_{i\alpha}n_{i\beta}}{\psi_0}=\frac{1}{4}$, implying 
$\lim_{|g|\to0}\expval{\hat D}{\psi_g}=0$. Therefore, we find $\lim_{|g|\to0}\partial_{|g|}\log{p_0}=-\frac{3}{2}N_{\text{site}}$, which indicates that $p_0$ decreases exponentially with $N_{\rm site}$ in the small-$|g|$ regime. 
On the other hand, for the BCS-like trial state, two colors form Cooper pairs while the third remains unpaired. This can lead to a larger expectation value of $\hat D$, as reflected in Eq.~\eqref{eq:bcs-d}.
In the opposite limit $|g|\to\infty$, the Gutzwiller wave function projects onto triply occupied states, giving $\lim_{|g|\to\infty}\expval{\hat D}{\psi_g}=\frac{3}{4}N_{\text{site}}$. Substituting this into Eq.~(\ref{eq:dp0-expression}), we obtain $\lim_{|g|\to\infty}\partial_{|g|}\log{p_0}=0$, implying that the success probability $p_0$ saturates in the large-$|g|$ limit for both choices of trial states.

Before proceeding further, we briefly review the algorithm for preparing trial states using a sequence of Givens rotation gates $\hat\G$ on a quantum computer, as described in Refs.~\cite{Wecker2015Solving,Jiang2018Quantum}.
For the Fermi-sea trial state $\ket{\psi_s}$,
the creation operators can be expressed as $\hat b_{j}^{\dagger}=\sum_{k}Q_{jk}\hat c_{k}^{\dagger}$, where $Q$ is an $N_f$ by $3N_{\text{site}}$ matrix whose row vectors correspond to the eigenvectors of the single-particle Hamiltonian $\hat K$.
Let $\ket{\psi_s}_c$ and $\ket{\psi_s}_b$ denote the many-body states constructed from the fermion creation operators $\hat c_{i}^{\dagger}$ and $\hat b_{j}^{\dagger}$, respectively.
The key idea is to find a sequence of Givens rotations $\G$ that transforms the projector matrix $Q^{\dagger}Q$ into a diagonal form. Suppose that this is achieved by transforming $Q$ to $VQ\G_1\G_2\dots\G_n$, where the basis transformation $V$ serves to eliminate the upper right elements of $Q$ \cite{Jiang2018Quantum}.
Then, according to \cite{Jiang2018Quantum}, the state $\ket{\psi_s}_c$ can be obtained via 
\begin{equation}\label{eq:get-psi0}
    \ket{\psi_s}_c = \hat\G_1\hat\G_2\dots\hat\G_n\ket{\psi_s}_b,
\end{equation}
up to an overall phase.
Both the Givens rotations $\G$ and the state $\ket{\psi_s}_b$ are easily prepared in the computational basis.
Therefore, by applying the reversed Givens rotation sequence to $\ket{\psi_s}_b$, we obtain the desired Fermi-sea trial state $\ket{\psi_s}_c$ according to Eq.~(\ref{eq:get-psi0}).

To prepare the BCS-like trial state defined in Eq.~\eqref{eq:bcs-trial}, we introduce the BCS mean-field Hamiltonian:
\begin{equation}
  \hat{H}_{\text{BCS}} = \hat K + \sum_{i}(\Delta c_{i1}^{\dagger}c_{i2}^{\dagger}+\text{H.c.}).
\end{equation}
The target BCS-like trial state corresponds to the ground state of $\hat{H}_{\text{BCS}}$ and can be constructed using a sequence of Givens-rotation gates, following similar procedures as in the case of the Fermi-sea state. Specifically, the Slater determinant for the color-3 fermion is prepared using the algorithm described above, while the pairing state between color-1 and color-2 fermions is generated using the method proposed in Ref.~\cite{Jiang2018Quantum}, which enables the preparation of arbitrary fermionic Gaussian states.

\begin{figure}[tb]
    \includegraphics[width=0.96\linewidth]{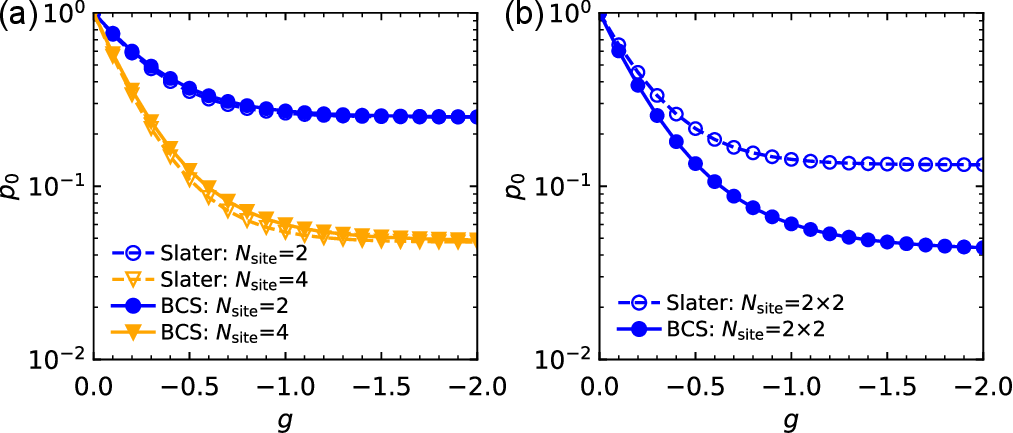}
    \caption{
      Success probability $p_0$ as a function of the variational parameter $g$ for (a) one-dimensional chains with open-boundary conditions and (b) a $2\times 2$ square lattice. Open symbols correspond to results using the Fermi-sea trial state, while solid symbols represent results using the BCS-like trial state with $\Delta/J=0.4$.
    }            
    \label{fig:p0}
\end{figure}

In our numerical simulations, we compare the performance of the Fermi-sea and BCS-like trial states using the first approach, which involves an exact summation over all auxiliary fields on small lattice sizes.
Figure~\ref{fig:p0} shows the success probability $p_0$ as a function of the variational parameter $g$ for different lattice geometries and sizes. 
The open symbols correspond to results obtained with the Fermi-sea trial state, while the solid symbols represent those using the BCS-like trial state.
As shown in Fig.~\ref{fig:p0}, the success probability decreases exponentially from $g=0$ and eventually saturates in the large $|g|$ limit for both types of trial states, consistent with the analytical behavior described in Eqs.~\eqref{eq:p0-expression} and \eqref{eq:dp0-expression}.
On one-dimensional chains, the BCS-like trial state exhibits a slightly slower decay of $p_0$ compared to the Fermi-sea case.  
In contrast, on a $2\times 2$ square lattice, the Fermi-sea trial state results in a slower decay, suggesting higher double occupancy, likely due to finite-size effects (see Sec.~\ref{sec:implemention-III} for further discussions).
Moreover, the BCS-like trial state introduces an additional variational parameter, $\Delta$, which adds complexity to the analysis.
For these reasons, we primarily focus on the Fermi-sea trial state in the remainder of this paper.

\subsection{Approach II: Importance sampling method} 
\label{sec:implementation-II}

The second approach aims to compute the expectation values of observables $\hat O$, which are central to the Gutzwiller variational calculation, without explicitly preparing the Gutzwiller wave function on a quantum computer. 
Using Eq.~\eqref{eq:gutzwiller-exp}, the expectation value with respect to the Gutzwiller wave function can be written as
\begin{equation}
    \expval{\hat O}{\psi_g} = \frac{\expval{e^{-g\hat D}\hat Oe^{-g\hat D}}{\psi_0}}{\expval{e^{-2g\hat D}}{\psi_0}}.
\end{equation}
After applying the HS transformation, both the numerator and the denominator involve a summation over auxiliary field configurations. This allows the expectation value to be reformulated as 
\begin{equation}\label{eq:mc_sum}
    \expval{\hat O}{\psi_g} = \sum_{\bm s} P_{\bm s} \langle \hat O\rangle_{\bm s},
\end{equation}
where $\bm s$ denotes a particular configuration of discrete auxiliary fields. 

The probability distribution $P_{\bm s}$ corresponding to configuration $\bm s$ is given by 
\begin{equation}
    P_{\bm s} = \frac{\expval{\prod_{i}e^{s_{i\alpha\beta,2}\lambda\iu\hat \sigma^y_{i\alpha\beta}}\prod_{j}e^{s_{j\zeta\eta,1}\lambda\iu\hat \sigma^y_{j\zeta\eta}}}{\psi_0}}
    {\sum_{\bm s'}\expval{\prod_{i}e^{s'_{i\alpha\beta,2}\lambda\iu\hat \sigma^y_{i\alpha\beta}}\prod_{j}e^{s'_{j\zeta\eta,1}\lambda\iu\hat \sigma^y_{j\zeta\eta}}}{\psi_0}},  
    \label{eq:Ps}
\end{equation}
and importantly $P_{\bm s}>0$ for our choices of initial trial states. Here, 
two sets of auxiliary fields, $\{s_{j\zeta\eta,1}\}$ 
and $\{s_{i\alpha\beta,2}\}$, are introduced by applying the HS transformation to $|\psi_g\rangle$ and $\langle\psi_g|$, respectively. 
The corresponding estimator $\langle \hat O\rangle_{\bm s}$ for the observable $\hat O$ is given by 
\begin{equation}\label{eq:mc_expval}
    \langle \hat O\rangle_{\bm s} = \frac{\expval{\prod_{i}e^{s_{i\alpha\beta,2}\lambda\iu\hat \sigma^y_{i\alpha\beta}}\hat O\prod_{j}e^{s_{j\zeta\eta,1}\lambda\iu\hat \sigma^y_{j\zeta\eta}}}{\psi_0}}
    {\expval{\prod_{i}e^{s_{i\alpha\beta,2}\lambda\iu\hat \sigma^y_{i\alpha\beta}}\prod_{j}e^{s_{j\zeta\eta,1}\lambda\iu\hat \sigma^y_{j\zeta\eta}}}{\psi_0}}.
\end{equation}
This expression can be interpreted as the matrix element corresponding to a component of the Gutzwiller wave function, i.e. the two factors appearing in Eqs.~\eqref{eq:HS-GutzF} and \eqref{eq:expD-decomposed}. 
For notational simplicity, we have omitted explicit product symbols such as $\prod_{\alpha<\beta}$ and $\prod_{\zeta<\eta}$ in Eqs.~(\ref{eq:Ps}) and (\ref{eq:mc_expval}), which denote products over color index pairs ($12$, $23$, and $13$).

Using Eq.~\eqref{eq:given-rotation}, the numerator of $P_{\bm s}$, as well as the denominator of $\langle \hat O\rangle_{\bm s}$, can be mapped--after applying the Jordan-Wigner transformation--as 
\begin{equation}\label{eq:numerator-P}
    \begin{aligned}
        &\expval{\prod_{i}e^{s_{i\alpha\beta,2}\lambda\iu\hat \sigma^y_{i\alpha\beta}}\prod_{j}e^{s_{j\zeta\eta,1}\lambda\iu\hat \sigma^y_{j\zeta\eta}}}{\psi_0} \\
        &\mapsto
        \expval{\prod_{i}\widehat\fG_{i_{\alpha}i_{\beta}}(s_{i\alpha\beta,2}\lambda)\prod_{j}\widehat\fG_{j_{\zeta}j_{\eta}}(s_{j\zeta\eta,1}\lambda)}{\psi_0}
     \\
        &:=\expval{\hat{\cal D}(\boldsymbol{s})}{\psi_0},
    \end{aligned}
\end{equation}
where we define $\hat{\cal D}(\boldsymbol{s})$ as the full sequence of fermionic Givens-rotation gates corresponding to auxiliary field configuration $\bm s$. Similarly, the numerator of $\langle \hat O\rangle_{\bm s}$ transforms as 
\begin{equation}\label{eq:numerator-H}
    \begin{aligned}
        &\expval{\prod_{i}e^{s_{i\alpha\beta,2}\lambda\iu\hat \sigma^y_{i\alpha\beta}}\hat O\prod_{j}e^{s_{j\zeta\eta,1}\lambda\iu\hat \sigma^y_{j\zeta\eta}}}{\psi_0} \\
        &\mapsto
        \expval{\prod_{i}\widehat\fG_{i_{\alpha}i_{\beta}}(s_{i\alpha\beta,2}\lambda)
        \hat O
        \prod_{j}\widehat\fG_{j_{\zeta}j_{\eta}}(s_{j\zeta\eta,1}\lambda)}{\psi_0}  \\
        & := \expval{\hat{\cal N}(\boldsymbol{s})}{\psi_0},
    \end{aligned}
\end{equation}
where $\hat{\cal N}(\boldsymbol{s})$ denotes the operator sequence acting on $|\psi_0\rangle$ in the presence of observable $\hat O$. 
Note that Eq.~\eqref{eq:numerator-H} reduces to Eq.~\eqref{eq:numerator-P} in the special case $\hat O=\hat I$. 

\begin{figure}[tb]
    \includegraphics[width=0.96\linewidth]{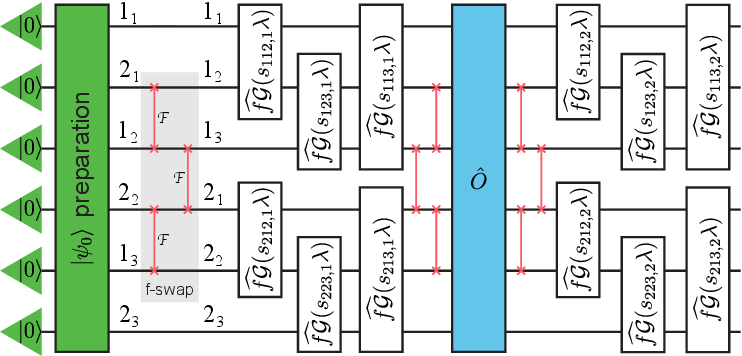}
    \caption{
        Quantum circuit for preparing the state $\hat{\cal N}(\boldsymbol{s}) |\psi_0\rangle$
        for $N_{\text{site}}=2$, corresponding to the numerator in Eq.(\ref{eq:mc_expval}). The shaded region represents a sequence of fermionic SWAP gates used to convert between color-uniform and color-alternating labelings.}
    \label{fig:Q-circuit-MC}
\end{figure}

The quantum circuit for preparing the state $\hat{\cal N}(\boldsymbol{s}) |\psi_0\rangle$ for the $N_{\text{site}}=2$ case, similar to that in Ref.~\cite{Seki2022Gutzwiller}, 
is shown in Fig.~\ref{fig:Q-circuit-MC}. 
Here, we introduce fermionic SWAP (f-SWAP) gates to map between color-uniform and color-alternating labelings.
The matrix element $\expval{\hat{\cal N}(\boldsymbol{s})}{\psi_0}$ in Eq.~(\ref{eq:numerator-H}) can be evaluated using the Hadamard test. 
This is done by introducing an ancillary qubit, modifying $\hat{\cal N}(\boldsymbol{s})$ into a controlled-$\hat{\cal N}(\boldsymbol{s})$ operator with control on the ancilla,  
and applying appropriate basis transformation, such as Hadamard gates, to the ancilla qubit before and after the controlled operation. Importantly, 
in the controlled-${\cal N}(\boldsymbol{s})$ operation, the f-SWAP gates need not be controlled by the ancilla qubit, since they appear symmetrically with their hermitian conjugates and cancel out in the expectation value. 
The matrix element $\expval{\hat{\cal D}(\boldsymbol{s})}{\psi_0}$ can be evaluated in the same manner by replacing $\hat{O}$ with the identity operator $\hat{I}$ in $\hat{\cal N}(\boldsymbol{s})$.

In this approach, the expectation value in Eq.~\eqref{eq:mc_sum} is evaluated via Monte Carlo sampling by generating a sequence of auxiliary field configurations $\bm s$ drawn from the probability distribution $P_{\bm s}$. 
To generate these configurations, we employ the Metropolis-Hastings algorithm. Starting from a current configuration $\bm s$, a candidate configuration $\bm s'$ is proposed stochastically according to simple update rules, such as local updates. 
A local update is defined as flipping a single auxiliary spin, $s_{i\alpha\beta,\tau}\to -s_{i\alpha\beta,\tau}$, within the current configuration. 
The acceptance probability for the proposed move $\bm s\to\bm s'$ is given by $p(\bm s\to\bm s') = \min(1,P_{\bm s'}/P_{\bm s})$. If the move is accepted, the candidate configuration $\bm s'$ is adopted for the next Monte Carlo step; otherwise, the current configuration $\bm s$ is retained. 
In our simulations, the auxiliary spins $s_{i\alpha\beta,\tau}$ are updated sequentially in the order $i=1,2,\dots N_{\text{site}}$, $\alpha\beta=12,23,13$, and $\tau=1,2$. A full cycle of such local updates constitutes one Monte Carlo sweep. To ensure convergence of the Metropolis algorithm to the equilibrium distribution, we discard the first $N_{\text{w}}$ sweeps as thermalization (or ``warmup") steps. Subsequently, measurements of observables, such as Eq.~\eqref{eq:mc_expval}, are performed after each Monte Carlo sweep. The total number of measurements is denoted by $N_{\text{m}}$.

\subsection{Results on numerical simulations} 
\label{sec:implemention-III}

\begin{figure}[tb]
    \includegraphics[width=0.96\linewidth]{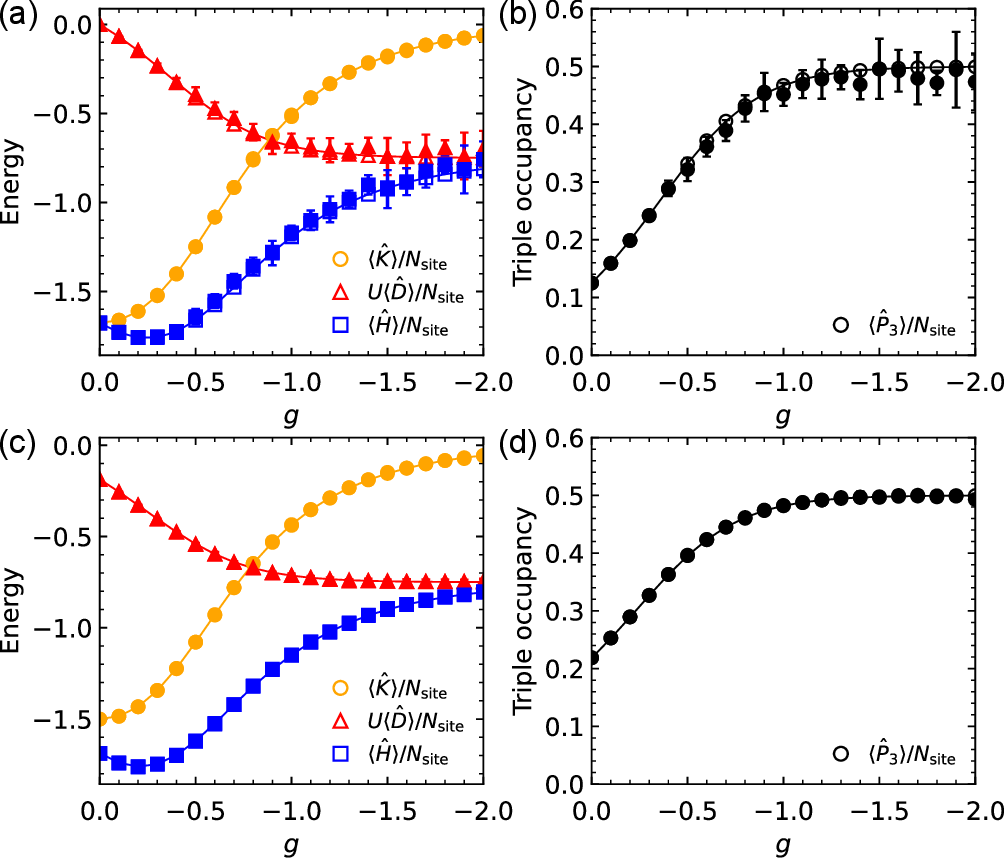}
    \caption{
      Expectation values of (a,c) the total energy, kinetic energy, and interaction energy, and (b,d) the triple occupancy per site as functions of $g$. Results are shown for (a,b) a 4-site one-dimensional chain with open boundary conditions and (c,d) a $2\times 2$ square lattice with $U=-1$. The Fermi-sea trial state at half filling is used. Open symbols represent results obtained by fully summing over all auxiliary field configurations (approach I; see Sec.~\ref{sec:implementation-I}), while solid symbols are from Monte Carlo sampling (approach II; see Sec.~\ref{sec:implementation-II}), with error bars indicating statistical errors. Solid lines correspond to the exact solutions. 
      Note that, due to the scale of the plots, results from approaches I and II are often indistinguishable. 
    }
    \label{fig:energy}
\end{figure}

To examine the two approaches for implementing the Gutzwiller operator in the attractive SU(3) Hubbard model, introduced in Sec.~\ref{sec:implementation-I} and Sec.~\ref{sec:implementation-II}, respectively, we perform numerical simulations of the quantum circuits shown in Fig.~\ref{fig:Q-circuit}(b) and Fig.~\ref{fig:Q-circuit-MC}. 
We evaluate the expectation values of several observables with respect to the Gutzwiller wave function, including the kinetic energy $\langle{\hat K}\rangle$, the interaction energy $U\langle{\hat D}\rangle$, the total energy $\langle{\hat H}\rangle =\langle{\hat K}\rangle + U\langle{\hat D}\rangle$, and the triple occupancy $\langle{\hat P_3}\rangle=\sum_i\langle{\hat n_{i1}\hat n_{i2}\hat n_{i3}}\rangle$, all as functions of $g$.
After applying the Jordan-Wigner transformation, the kinetic and interaction energies are expressed in Eqs.~\eqref{eq:JW-K} and \eqref{eq:JW-D}, respectively. The triple occupancy operator is transformed as 
\begin{equation}\label{eq:JW-P3}
    \begin{aligned}
        \hat P_3 &\mapsto \frac{1}{8}\sum_{i=1}^{N_{\text{site}}}(\hat I-\hat Z_{i_{1}})(\hat I-\hat Z_{i_{2}})(\hat I-\hat Z_{i_{3}}) \\
          &
        =\frac{1}{2}\sum_{i=1}^{N_{\text{site}}}(\hat I - \CCZ_i), 
    \end{aligned}
\end{equation}
where $\CCZ_i$ denotes the controlled-CZ operator acting on the $i_1$th, $i_2$th, and $i_3$th qubits. 
Note that the first line of Eq.~(\ref{eq:JW-P3}) corresponds to a sum of projectors onto the triply occupied states: $\sum_{i=1}^{N_{\rm site}} |111\rangle_{ii}\langle 111|$, where $|111\rangle_{ii}\langle 111| := \prod_{\alpha=1}^{3}|1\rangle_{i_\alpha i_\alpha} \langle 1|$. 
For the importance sampling approach, the Monte Carlo simulations are performed without encountering the phase problem. The results are averaged over 
16 independent simulations, 
each using $N_{\text{w}}=2000$ warmup sweeps and $N_{\text{m}}=2000$ measurement sweeps.

Figure~\ref{fig:energy} presents the results obtained using the Fermi-sea state at half filling as the trial state $\ket{\psi_0}$, for both a 4-site one-dimensional chain with open boundary conditions [Figs.~\ref{fig:energy}(a) and \ref{fig:energy}(b)] and a $2\times2$ square lattice [Figs.~\ref{fig:energy}(c) and \ref{fig:energy}(d)]. 
Open symbols correspond to results obtained by fully summing over all auxiliary field configurations (approach I), while solid symbols denote results obtained using the Monte Carlo importance sampling method (approach II). 
For reference, the corresponding exact solutions are also shown by solid lines. The results clearly demonstrate that both approaches accurately reproduce the exact values (within the statistical errors in the case of approach II).

As shown in Figs.~\ref{fig:energy}(a) and \ref{fig:energy}(c), the total energy initially decreases with increasing $|g|$, reaches a minimum (corresponding to the ground state energy), and then rises again as $|g|$ becomes larger. 
In Fig.~\ref{fig:energy}(c), the interaction energy at $g=0$ is nonzero, which is attributed to finite-size effects: the exact solution shows that the interaction energy at $g=0$ becomes vanishingly small for systems with $N_{\text{site}}\geqslant4\times 4$. 
In Figs.~\ref{fig:energy}(b) and \ref{fig:energy}(d), the triple occupancy increases monotonically with $|g|$, approaching the asymptotic value $0.5$, which signals the formation of trionic states.
Notably, in Figs.~\ref{fig:energy}(a) and \ref{fig:energy}(b), while the importance sampling method (approach II) successfully reproduce the exact results, the statistical errors associated with the interaction energy and the triple occupancy are significantly larger than those for the kinetic energy. In contrast, the corresponding statistical errors in Figs.~\ref{fig:energy}(c) and \ref{fig:energy}(d) are smaller than the symbol sizes, making them practically invisible.  
Since the $N_{\text{site}}=2\times 2$ square lattice is equivalent to a 4-site one-dimensional chain with periodic boundary conditions, this observation suggests that the relatively large statistical errors in the open-boundary chain may arise from boundary effects.

\subsection{Results on quantum computer} 
\label{sec:implementation-IV}

While the proposed approaches accurately reproduce the exact results on classical simulations, we now report experimental implementations of our methods for the two-site attractive SU(3) Hubbard model, using both quantum simulators and a real quantum device.
Specifically, we employ the Quantum Information Software Kit (Qiskit)~\cite{Qiskit,qiskit2024} to construct the quantum circuits and perform simulations using the ideal simulator (aer\_simulator) and the noise model corresponding to IBM Torino device (fake\_torino). 
Furthermore, we execute the circuits on the Quantinuum H1-1 trapped-ion quantum computer using the $\rm t|ket\rangle$ compiler~\cite{tket}.

The experiments on the Quantinuum H1-1 system were carried out in May and June 2024.
At the time, the H1-1 system featured $20$ qubits and natively supported single-qubit rotation gates as well as two-qubit ZZ phase gates, defined as ${\rm ZZPhase}(\alpha) = e^{-\frac{1}{2} i \pi \alpha \hat{Z}_i \hat{Z}_j}$, 
where $\alpha$ is real-valued parameter. 
Importantly, these two-qubit gates could be applied between any pair of qubits. 
The average gate infidelity was approximately $0.002\%$ for single-qubit gates and $0.1\%$ for two-qubit gates. State preparation and measurement (SPAM) errors were around $0.3\%$. 
Further hardware specifications are available in 
Ref.~\cite{h1-1}.

We begin by discussing the practical feasibility of the two approaches for the two-site attractive SU(3) Hubbard model.
The importance sampling method (approach II) employs simplified quantum circuits that do not require $3N_{\text{site}}$ ancillary qubits to evaluate the numerators and denominators in Eqs.~\eqref{eq:mc_sum}--\eqref{eq:mc_expval}. In Ref.~\cite{Seki2022Gutzwiller}, the total number of auxiliary field configurations was only 16, allowing the summation to be carried out manually. 
In contrast, in our model, the number of configurations grows significantly , with a total of $2^{12}=4096$ auxiliary fields. Furthermore, real quantum hardware impose additional constraints, such as limited qubit connectivity. As a result, implementing the controlled-$\widehat{\fG}$ operators required for the Hadamard test may demand either the generation of a GHZ state across $3N_{\text{site}}$ ancillary qubits~\cite{Summer2024}, or a carefully engineered sequence of SWAP gates when using a single ancilla qubit.

On the other hand, the first approach (approach I) performs the summation over all auxiliary field configurations by postselecting the state where all $3N_{\text{site}}$ ancillary qubits are measured in $\ket{0\dots0}$, as described in Sec.~\ref{sec:implementation-I}. Although probabilistic, this method achieves a success probability 0.1 for the two-site model [see Fig.~\ref{fig:p0}(a)], making it a practically viable strategy. 
Accordingly, we demonstrate this first approach by explicitly preparing the Gutzwiller wave function on a quantum computer using a 12-qubit experiment.

\begin{figure*}[tbh]
    \includegraphics[width=0.98\linewidth]{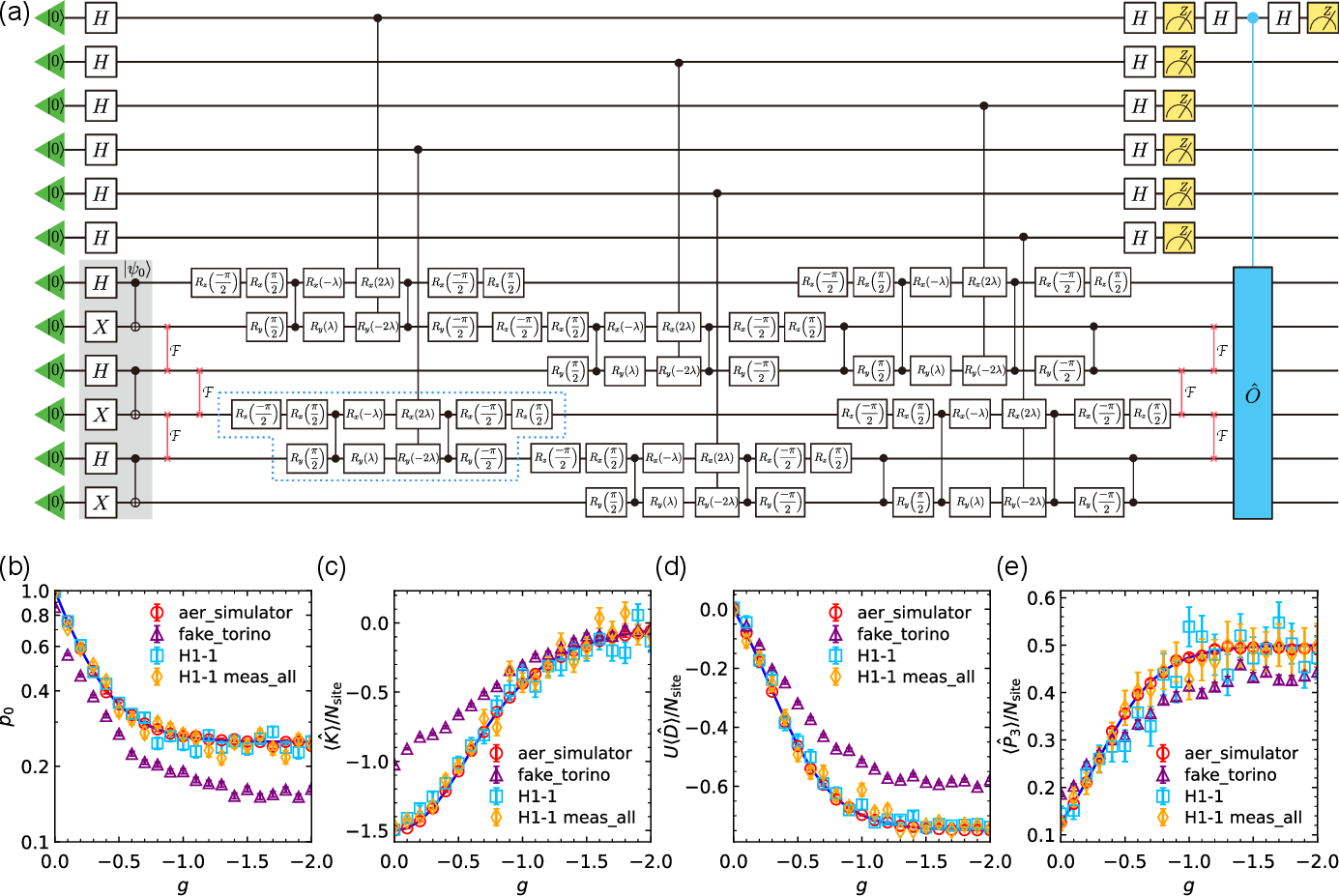}
    \caption{
        (a) Quantum circuit for the Hadamard test used to evaluate $\langle{ \hat O }\rangle$ for the two-site attractive SU(3) Hubbard model. 
        (b) Success probability, (c) kinetic energy per site, (d) interaction energy per site, and (e) triple occupancy per site as functions of the Gutzwiller parameter $g$. 
        Numerical results from ideal (aer\_simulator) and noisy (fake\_torino) simulations are obtained using $N_{\rm shots}=8192$ shots per data point, where the ancillary qubit is measured in the computational basis to estimate the expectation values. For the experiments on the Quantinuum H1-1 system (H1-1 and H1-1 meas\_all), the number of shots is fixed at $N_{\rm shots}=256$ per data point. The model parameter is set to $U=-1$ for panels (b)--(e). For comparison, the exact solutions are also shown by solid lines. 
    }
    \label{fig:Q-circuit-qsimu-test}
\end{figure*}

Figure~\ref{fig:Q-circuit-qsimu-test}(a) shows the quantum circuit used to prepare the Gutzwiller wave function and evaluate the expectation value $\langle{\hat O}\rangle$ of a given observable $\hat O$.
In the two-site case, the Pauli $X$, Hadamard, and CNOT gates are employed to prepare the Fermi-sea trial state for each fermion color, which corresponds to a Bell state~\cite{Seki2022Gutzwiller}, as indicated by the gray-shaded box in Fig.~\ref{fig:Q-circuit-qsimu-test}(a). 
The blue dashed box highlights a sequence of fermionic Givens and controlled-Givens rotation gates acting between two fermion colors. Each of these gates is decomposed into a set of eight single-qubit rotations, two CZ gates, and two controlled-$R$ (C$R$) gates.
Following the Hadamard-test-like subcircuits, measurements are performed on the six ancillary qubits, postselecting the state $\ket{0}^{\otimes 6}$. Upon successful postselection, the register qubits are projected into the Gutzwiller wave function.
It is worth noting that, in principle, a single ancillary qubit could be reused $3N_{\text{site}}$ times to prepare the Gutzwiller wave function via repeated postselection. However, in our experiments, we use $3N_{\text{site}}$ ancillary qubits simultaneously in order to reduce SPAM errors.

It is important to note that for quantum devices with limited connectivity, the circuit shown in Fig.~\ref{fig:Q-circuit-qsimu-test}(a) is better suited to quantum hardware with a 2D square-lattice architecture, 
as opposed to heavy-hex lattice designs.
Nonetheless, we can utilize the transpilation tools provided by Qiskit to automatically adapt our circuit to the constraints of the fake\_torino backend, at the cost of introducing additional SWAP gates.
By contrast, the Quantinuum H1-1 system offers full all-to-all connectivity, enabling direct compilation of the circuit into its native gate set without requiring any additional SWAP gates.

Alternatively, instead of using the Hadamard test, one can apply a direct measurement approach to evaluate expectation values from the register qubits. 
Specifically, the kinetic energy (interaction energy) can be obtained by measuring the register qubits in the $X$ and $Y$ bases (in the $Z$ basis), as described in Eqs.~\eqref{eq:JW-K} and \eqref{eq:JW-D}.
Note that the Jordan-Wigner string appearing in Eq.~\eqref{eq:JW-K} reduces to the identity operator because the qubits are measured using the color-uniform labeling scheme at the time of measurement. 
The corresponding results obtained using this direct measurement strategy on the H1-1 system are referred to as ``H1-1 meas\_all'' for comparison.

In Sec.~\ref{sec:implemention-III}, we computed the expectation values of observables with respect to the Gutzwiller wave function using numerical simulations, where the number of measurement shots was implicitly assumed to be infinity. 
In practice, however, the number of shots, $N_{\text{shots}}$, is finite and thus statistical uncertainty arises due to shot noise~\cite{seki2024simulating}.
In the following, we describe how the success probability and the expectation values of observables are evaluated under a finite number of shots.

First, we describe how the success probability is evaluated. 
For convenience, we define a projection operator $\hat{\cal P} := (|0\rangle \langle 0|)^{\otimes 6} \otimes \hat{I}^{\otimes 6}$, 
where the first six qubits correspond to the ancillary register. 
Let ${\cal P} \in \{0, 1\}$ denotes the eigenvalues of $\hat{\cal P}$. 
The success probability is then estimated as 
\begin{equation}
  \check{p}_0 = \mu \left[{\cal P} \right] := \frac{1}{N_{\text{shots}}} \sum_{c=1}^{N_{\text{shots}}} {\cal P}_{c},
\end{equation}
where the check accent ($\check{p}_0$) indicates that this is an estimator of the success probability $p_0$ in Eq.~\eqref{eq:p0-expression}. Here, $\mu[\cdot]$ denotes the sample mean, and ${\cal P}_{c}=\{0, 1\}$ is the projective measurement outcome for shot $c$: ${\cal P}_c=1$ if the six ancillary qubits are measured in the state $|000000\rangle$, and ${\cal P}_c=0$ otherwise.
The sample variance of the estimator is given by 
\begin{equation}
  \mathrm{var}[{\cal P}] = \mu[{\cal P}^2] - (\mu[{\cal P}])^2 = \check{p}_0 - \check{p}_0^2,
  \label{eq:varp0}
\end{equation}
where we have used the identity ${\cal P}^2 = {\cal P}$.  
The standard error due to shot noise is then give by  
$\epsilon_{\mu[{\cal P}]}=\sqrt{\mathrm{var}[{\cal P}]/N_{\text{shots}}}$. 

Next, we explain how observables $\hat{O} \in
\{\hat{X}_{i_{\alpha}} \hat{X}_{j_{\alpha}},\hat{Y}_{i_{\alpha}} \hat{Y}_{j_{\alpha}},\hat{Z}_{i_{\alpha}} \hat{Z}_{i_{\beta}}\}$
 are evaluated using the Hadamard test.
To this end, we introduce a Pauli operator acting on an additional ancillary qubit,
 $\hat{\cal Z}=\hat{Z} \otimes \hat{I}^{\otimes 12}$, 
and let ${\cal Z} \in \{+1,-1\}$ denote the eigenvalues of $\hat{\cal Z}$. 
The expectation value of the observable is then estimated via the Hadamard test as 
\begin{equation}
  \langle \hat{O} \rangle =
  \mu'[{\cal Z}] := \frac{1}{N_{\text{shots}}'}\sum_{c=1}^{N_{\text{shots}}'}{\cal Z}_c,
\end{equation}
where ${\cal Z}_c\in\{+1,-1\}$ represents the outcome of measuring the ancillary qubits in the states $|0000000\rangle$ and $|0000001\rangle$ for the $c$th shot. 
Here, we have introduced the effective number of shots
\begin{equation}
  N_{\text{shots}}'=p_0 N_{\text{shots}},
  \label{eq:Nshots_eff}
\end{equation}
which corresponds to the number of shots in which the Gutzwiller wave function is successfully prepared on the register qubits. 
Note that first six zeros in the bitstring outcomes $|0000000\rangle$ and $|0000001\rangle$ indicate the successful preparation of the Gutzwiller wave function, while the last bit (0 or 1) corresponds to the measurement outcome of the additional ancillary qubit in the computational basis. 
The associated variance is given by 
\begin{equation}
  \mathrm{var}'[{\cal Z}] =
  \mu'[{\cal Z}^2] - (\mu'[{\cal Z}])^2 
  =1-\langle \hat{O} \rangle^2,
\end{equation}
where ${\cal Z}^2=1$ is used.
The standard error of the sample mean is then calculated as $\epsilon_{\mu[\cal Z]} = \sqrt{\mathrm{var}'[{\cal Z}]/N_{\text{shots}}'}$. 
In our experiments, we reuse the first ancillary qubit for the Hadamard test, as shown in Fig.~\ref{fig:Q-circuit-qsimu-test}(a). 

Figures~\ref{fig:Q-circuit-qsimu-test}(b)--\ref{fig:Q-circuit-qsimu-test}(e) summarize the outcomes of the quantum circuits executed on the ideal simulator, noisy simulator, and the Quantinuum H1-1 system, in comparison with the exact results.
While the results obtained from the ideal simulator show excellent agreement with the exact solutions, the raw data from the fake Torino backend exhibit noticeable discrepancies. 
To mitigate such systematic errors, the phase-and-scale error mitigation technique is commonly employed~\cite{Seki2022Gutzwiller,chiesa2019quantum,Francis2020Quantum}.
In contrast, the raw data from the H1-1 system closely match the exact results within the error bars, demonstrating its ability to reliably prepare the Gutzwiller wave function and to solve the Gutzwiller problem.
As noted above, the average single- and two-qubit gate infidelities in the Quantinuum H1-1 system are about $0.002\%$ and $0.1\%$, respectively~\cite{h1-1}, whereas for the fake Torino backend, the corresponding values are around $0.04\%$ (single-qubit) and $0.6\%$ (two-qubit)~\cite{fake-torino}.
These observations suggest that the good agreement between the raw experimental results and the exact solutions can be attributed not only to the all-to-all connectivity of the Quantinuum system, but also to its exceptionally low gate infidelities.

On another note, as indicated by the second line of Eq.~\eqref{eq:JW-P3}, the circuit for the Hadamard test used to evaluate $\langle\hat P_3\rangle$ involves the implementation of a controlled-controlled-CZ (CC-CZ) gate, which is compiled to 50 ZZPhase gates. This is in contrast to 39 ZZPhase gates required for evaluating other observables such as $\langle{\hat K}\rangle$ and $U\langle{\hat D}\rangle$~\cite{tket}, and it leads to noticeable deviations from the exact results due to increased gate infidelities.
Alternatively, as indicated by the first line of Eq.~\eqref{eq:JW-P3}, the direct measurement strategy requires only a final measurement of all the qubits in the computational basis and uses 37 ZZPhase gates for each observable listed above. This approach yields comparatively more accurate values of $\langle\hat P_3\rangle$, as shown in \ref{fig:Q-circuit-qsimu-test}(e). 
In this sense, the Hadamard test does not offer a clear advantage over direct measurement for the observables considered here, even though it generally requires fewer measurements per circuit. 
However, the Hadamard test becomes essential when evaluating dynamical quantities such as dynamical correlation functions~\cite{chiesa2019quantum}, out-of-time-ordered correlation functions~\cite{Mi2021}, and Loschmidt amplitudes~\cite{Summer2024,seki2024simulating}, as these involve expectation values of unitary but non-Hermitian operators.

Moving forward, we analyze the scalability of the quantum circuit in terms of the number of CNOT gates required to generate the Gutzwiller wave function for larger lattice systems.
When implementing the Gutzwiller operator, the basic structure involves Hadamard-test-like variants using $\widehat\fG$ and controlled-$\widehat\fG$ gates, where the gate decompositions shown in Figs.~\ref{fig:Q-rotation}(a) and \ref{fig:Q-rotation}(b) are employed. This results in the addition of 8 CZ, 3 C$R_x$, and 3 C$R_y$ gates per site. 
The CNOT gate count required to decompose each of these gates is as follows: 1 for CZ, 2 for C$R_x$, 2 for C$R_y$, and 3 for SWAP gate~\cite{Seki2022Gutzwiller}.
Therefore, for a lattice system with $N_{\text{site}}$ sites, the Gutzwiller operator requires a total of $20N_{\text{site}}$ CNOT gates.
In addition, initializing the Fermi-sea trial state requires at most $\frac{3}{4}N_{\text{site}}^2$ Givens rotations~\cite{Jiang2018Quantum}, which translates to $\frac{3}{2}N_{\text{site}}^2$ CNOT gates. This is comparable to the $6N_{\text{site}}(N_{\text{site}}-1)$ CNOT gates required for the f-SWAP operations that transform between color-uniform and color-alternating labelings.
In total, the estimated CNOT gate counts for the $6N_{\text{site}}$-qubit quantum circuit is given by $\frac{15}{2}N_{\text{site}}^2+14N_{\text{site}}$.

Lastly, let us provide a rough estimate of the quantum resources. 
To be specific, we estimate the quantum resources by accounting for error mitigation through division by the circuit fidelity: $\langle \hat{O} \rangle_{\rm mit} = \langle \hat{O} \rangle_{\rm raw}f^{-1}$, where $\langle \hat{O} \rangle_{\rm raw}$ and $\langle \hat{O} \rangle_{\rm mit}$ are the raw and mitigated expectation values of $\hat{O}$, respectively, and $\hat{O}$ is assumed to be traceless. Here, $f$ denotes the circuit fidelity~\cite{shinjo2024}.
Let $\delta O_{\rm raw}$ and $\delta O_{\rm mit}$ be the statistical uncertainties of the raw and mitigated expectation values, respectively.
Given that $\delta O_{\rm raw} \sim N_{\text{shots}}^{\prime -1/2}$, it follows that $\delta O_{\rm mit}  \sim N_{\text{shots}}^{\prime -1/2}f^{-1}$, where $N_{\text{shots}}'$ is the effective number of shots defined in Eq.~(\ref{eq:Nshots_eff}). 
To obtain the mitigated expectation values with a target statistical error $\delta O_{\rm mit} \sim \epsilon$, the required number of shots scales as $N_{\text{shots}} \sim O(p_0^{-1}f^{-2}\epsilon^{-2})$.  
Concretely, let us consider a system of $N_{\text{site}}=16$. 
The dimension of the full Hilbert space for the SU(3) fermion model is $2^{48}$, and the linear-combination-of-unitaries circuit involves $96$ qubits and approximately $2144$ two-qubit gates.  
Assuming that the dominant source of error arises from the two-qubit gates, with an average gate infidelity of $0.1\%$, the overall circuit fidelity, in the worst case, can be estimated as $f=0.999^{2144}\approx0.1$. 
Note that, in general, the effective number of two-qubit gates contributing to the circuit fidelity depends on both the circuit structure and the observable, and is often smaller than the total number of two-qubit gates within the causal cone~\cite{Kechedzhi_2024}.
Therefore, the above estimate of $f\approx 0.1$ should be regarded as a worst-case scenario.
It should be noted that $p_0$ decreases exponentially with $N_{\text{site}}$, potentially rendering experiments impractical for larger systems. 
Nonetheless, as suggested by Eq.~\eqref{eq:dp0-expression}, using a correlated initial state $\ket{\psi_0}$ that includes multiple occupancy can lead to a slower decay and, hence, a larger success probability. This improvement may enhance the feasibility of preparing the Gutzwiller wave function on a quantum computer for larger lattice systems.


\section{Conclusion and discussion}\label{sec:conc}

In summary, we have generalized a quantum-classical hybrid scheme for preparing the Gutzwiller wave function to attractively interacting SU(3) fermions.
For SU(3) fermions, the nonunitary Gutzwiller operator is expressed as a linear combination of fermionic Givens rotations between different fermion colors, derived by introducing discrete auxiliary fields through a variant of the Hubbard-Stratonovich transformation tailored to the attractive SU(3) Hubbard interaction.
We have reformulated two complementary approaches for summing over the auxiliary field configurations. The first is a probabilistic method that uses ancillary qubits in a linear-combination-of-unitaries circuit to probabilistically prepare the Gutzwiller wave function on a quantum computer. 
The second is a stochastic method based on importance sampling, which evaluates the expectation values with respect to the Gutzwiller wave function without preparing the wave function itself on the quantum hardware.  
We tested both approaches on small systems and computed physical observables, including the energy and triple occupancy, as functions of the Gutzwiller variational parameter. 
Furthermore, we experimentally demonstrated the first approach on a trapped-ion quantum computer for the two-site attractive SU(3) Hubbard model, obtaining success probabilities and energy values that quantitatively agree with the exact results.

The present experimental results demonstrate the capability of currently available quantum computers to probabilistically prepare the Gutzwiller wave function for attractive SU(3) fermions using the first approach. 
We note that this approach can be readily extended to systems away from half filling, without requiring significant modifications to the quantum circuit. 
Although we have not yet reached classically intractable lattice sizes, we expect that the proposed algorithm will be useful for preparing the Gutzwiller wave function for attractive SU(3) fermions on a quantum computer, particularly when initial states with  higher multiple occupancy are employed. 
Future extensions of the present experiments to larger lattices, aiming to investigate pairing and trion formations in attractive SU(3) fermion models, will be a promising direction of research. 


\section*{Acknowlegments}

A part of the numerical simulations was performed using the HOKUSAI supercomputer at RIKEN (Project ID RB240003).
A portion of this work is based on results obtained from project JPNP20017, supported by the New Energy and Industrial Technology Development Organization (NEDO). 
This study was also supported by JSPS KAKENHI 
Grants 
No. JP21H04446 and
No. JP22K03520.
We are additionally grateful for funding from JST COI-NEXT (Grant No. JPMJPF2221) and the Program for Promoting Research on the Supercomputer Fugaku (Grant No. MXP1020230411) by MEXT, Japan.  
We also acknowledge support from the UTokyo Quantum Initiative, 
the RIKEN TRIP project (RIKEN Quantum), and
the COE research grant in computational science from Hyogo Prefecture and Kobe City through the Foundation for Computational Science.

%

\end{document}